\documentclass[conference]{IEEEtran}
\IEEEoverridecommandlockouts
\usepackage{cite}
\usepackage{siunitx}

\usepackage{amsmath,amssymb,amsfonts}
\usepackage{algorithmic}
\usepackage{graphicx}
\usepackage{textcomp}
\usepackage{xcolor}
\usepackage{caption}
\usepackage{subcaption}
\def\BibTeX{{\rm B\kern-.05em{\sc i\kern-.025em b}\kern-.08em
    T\kern-.1667em\lower.7ex\hbox{E}\kern-.125emX}}

\usepackage{colortbl}

\newcommand{\Ac}{\mathcal{A}}

\newcommand{\Pc}{\mathcal{P}}

\newcommand{\Tc}{\mathcal{T}}

\usepackage{multirow}
\usepackage{url}            

\begin{document}

\title{Improving Perceptual Quality, Intelligibility, and Acoustics on VoIP Platforms}

\author{

\IEEEauthorblockN{
Joseph Konan\IEEEauthorrefmark{1}\IEEEauthorrefmark{2},
Ojas Bhargave\IEEEauthorrefmark{2},
Shikhar Agnihotri\IEEEauthorrefmark{2},
Hojeong Lee\IEEEauthorrefmark{2},
Ankit Shah\IEEEauthorrefmark{2}, \\
Shuo Han\IEEEauthorrefmark{2},
Yunyang Zeng\IEEEauthorrefmark{2},
Amanda Shu\IEEEauthorrefmark{2},
Haohui Liu\IEEEauthorrefmark{2},
Xuankai Chang\IEEEauthorrefmark{2}, \\
Hamza Khalid\IEEEauthorrefmark{2},
Minseon Gwak\IEEEauthorrefmark{2},
Kawon Lee\IEEEauthorrefmark{2},
Minjeong Kim\IEEEauthorrefmark{2},
Bhiksha Raj\IEEEauthorrefmark{2}}

\IEEEauthorblockA{
\IEEEauthorrefmark{1}KonanAI, Pittsburgh, United States, konan@konanai.com.  \\
\IEEEauthorrefmark{2}Carnegie Mellon University, Pittsburgh, \\
\{
jkonan,
obhargav,
sagnihot,
hojeongl,
aps1,
shuohan,
yunyangz,
amshu,
haohuil,
xuankaic,  \\
hkhalid,
mgwak,
kawonl,
minjeong,
bhikshar
\}@andrew.cmu.edu
}
}

\maketitle

\begin{abstract}
In this paper, we present a method for fine-tuning models trained on the Deep Noise Suppression (DNS) 2020 Challenge to improve their performance on Voice over Internet Protocol (VoIP) applications. Our approach involves adapting the DNS 2020 models to the specific acoustic characteristics of VoIP communications, which includes distortion and artifacts caused by compression, transmission, and platform-specific processing. To this end, we propose a multi-task learning framework for VoIP-DNS that jointly optimizes noise suppression and VoIP-specific acoustics for speech enhancement. We evaluate our approach on a diverse VoIP scenarios and show that it outperforms both industry performance and state-of-the-art methods for speech enhancement on VoIP applications. Our results demonstrate the potential of models trained on DNS-2020 to be improved and tailored to different VoIP platforms using VoIP-DNS, whose findings have important applications in areas such as speech recognition, voice assistants, and telecommunication.
\end{abstract}


\section{Introduction}

Voice over Internet Protocol (VoIP) has become a ubiquitous technology for real-time voice communications over the internet. However, speakers often call from environments with background noise that degrades perceptual quality and impairs intelligibility of the communication. To address this problem, denoising and speech enhancement has emerged as promising technique for improving the quality of speech signals in noisy environments. The Deep Noise Suppression (DNS) Challenge \cite{reddy2020interspeech}, hosted by InterSpeech 2020, developed a systematic training and evaluation methodology to improve state-of-the-art denoising and speech enhancement model performance on a diverse set of real-world scenarios with background noise.

Despite the success of models trained on the DNS-2020 Challenge, their performance on VoIP applications remains limited due to a lack of optimization of characteristics specific to VoIP platforms. VoIP audio is typically transmitted over low-bandwidth networks, compressed using lossy codecs, and subject to non-uniform sampling rates, which can introduce additional distortion and artifacts. To enable analysis and optimization in this research area, the VoIP Deep Noise Suppression (VoIP-DNS) dataset was developed as an extension to DNS-2020 that provides training and testing data for VoIP applications: Zoom and Google Meet.
Using a controlled experiment design, the VoIP-DNS dataset was created using consistent systems, hardware, network, and data collection procedures necessary for reproducible research. 

In this paper, we present a method for fine-tuning two state-of-the-art DNS-2020 models, Demucs \cite{defossez2020real} and FullSubNet\cite{hao2021fullsubnet}, on the VoIP-DNS dataset for improved performance on VoIP applications. Our approach involves adapting the models to acoustic variations present in VoIP audio streams, using a temporal acoustic parameter loss (TAPLoss)\cite{taploss} to optimize fine-grain speech characteristics. We evaluate our approach on two VoIP platforms, Zoom and Google Meets, in the context of cloud and cellular phone applications. Our results demonstrate the potential of fine-tuning Demucs and FullSubNet to surpass industry performance and achieve state-of-the-art on VoIP-DNS by improving the quality of voice communication on VoIP platforms, which has important applications the areas of speech recognition, voice assistants, and telecommunication.

\begin{figure*}[!htb]
    \centering
    \begin{subfigure}{0.5\textwidth}
        \includegraphics[width=\linewidth,height=6cm]{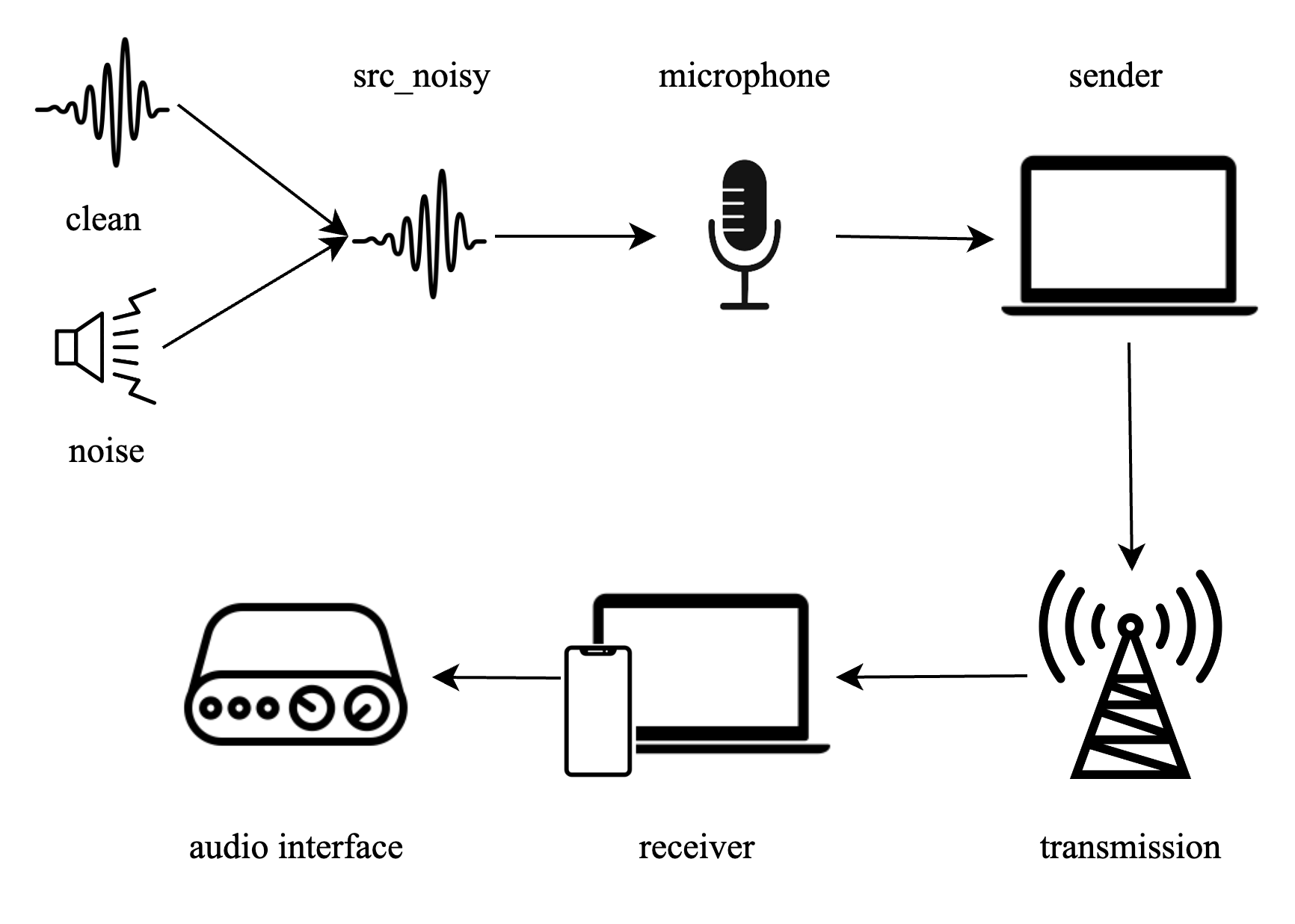}
        \caption{End-to-end Data Process Diagram, From Synthesis To Recording. }
        \label{fig:transmission_diagram}
    \end{subfigure}%
    \begin{subfigure}{0.5\textwidth}
        \includegraphics[width=\linewidth]{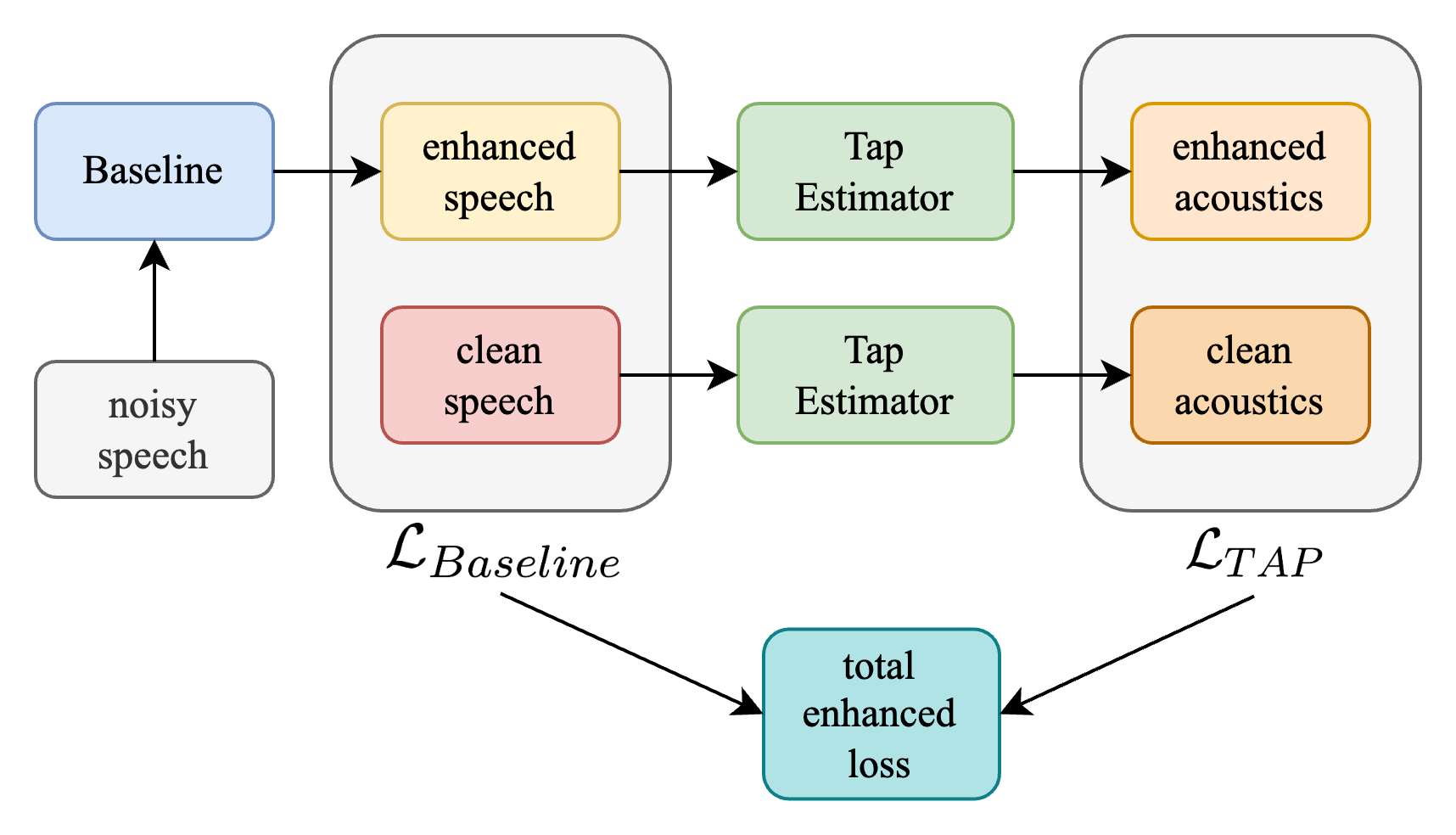}
        \caption{The training workflow for Demucs and FullSubNet}
        \label{fig:train1}
    \end{subfigure}
    \caption{Data Synthesis and the Training workflow}
    \label{fig:my_label}
\end{figure*}

\section{Experimental Setup}

The VoIP-DNS dataset\cite{voip} contains 1,200 thirty-second training pairs and 150 ten-second testing pairs of source clean speech and source noisy speech. The source pairs for the VoIP-DNS dataset are obtained using the same training synthesis procedure and test set provided by DNS-2020. The training synthesis procedure involves taking 30-second clean speech samples from LibriSpeech and 30-second noise signals from FreeSound\cite{fonseca2017freesound} or AudioSet\cite{gemmeke2017audio_audioset}, then mixing the two to create a 30-second noisy speech sample. After the source noisy speech is passed to the VoIP platform via a virtual microphone, it is transmitted to a receiving device. Finally, the relayed audio is obtained via an audio interface. If the receiver is a cellular phone, then the audio interface is managed by the VoIP-DNS technician; otherwise, if the receiver is on the cloud, then the audio interface is managed by the VoIP platform.

This paper focuses on optimizing VoIP-DNS to improve speech quality on cloud and cellular phones using VoIP data from Zoom and Google Meet. On each platform, we consider their use cases with denoising set on or off. Due to differences in terminology across platforms, disambiguation is required. For Zoom, "denoising off" is considered "Zoom with low denoising selected," while "denoising on" is considered "Zoom with auto denoising selected." For Google Meet, "denoising off" is considered "Google Meet with denoising off selected," while "denoising on" is considered "Google Meet with denoising on selected." Therefore, industrial denoising is defined by the platform's performance with denoising on.

The VoIP-DNS data collection procedure captures the unique characteristics of VoIP audio streams, which are subject to low bandwidth, compression, non-uniform sampling rates, and additional platform-specific processing. The VoIP-DNS dataset was designed by the authors of this work and is publicly available with a meticulously documented experimental design to ensure reproducibility.

\section{Models and Training}

In this study, we evaluate the denoising performance of two state-of-the-art deep learning models, Demucs and FullSubNet, on VoIP audio streams. These models were chosen due to their top performance in the Deep Noise Suppression (DNS) 2020 Challenge and their availability in terms of public source code, learning procedures, and final checkpoints. Furthermore, Demucs operates in the time-domain, while FullSubNet operates in the time-frequency domain, allowing us to contrast modalities with a consistent reproducible experimental setup.

To train the models, we use the noisy speech relays and the clean speech sources from VoIP-DNS. Each training iteration begins by sampling a noisy speech signal and its corresponding clean speech source. The noisy speech signal is passed through the base model to generate enhanced speech signals. To optimize the fine-grained acoustic characteristics of the enhanced speech signals, a temporal acoustic parameter (TAP) estimator is used to derive enhanced acoustic features from the enhanced speech signals. Similarly, clean acoustic features are derived from the clean speech source using the TAP estimator.

The base loss function is defined by the base model and minimizes the divergence between the enhanced and clean speech signals in either the time-domain (for Demucs) or time-frequency domain (for FullSubNet). To further improve the performance of the models on VoIP audio streams, we propose a novel acoustic loss function, which minimizes the L1 divergence between the enhanced and clean acoustic features. Our experimental results show that incorporating the acoustic loss function leads to significant improvements in the denoising performance of the models on VoIP audio streams.

In addition to evaluating the performance of Demucs and FullSubNet, we compare their performance with industrial denoising models used by popular VoIP platforms such as Zoom and Google Meet. While these industrial denoising models are not publicly available for inspection, they can be queried through their platform using the VoIP-DNS dataset. This enables a direct comparison between the performance of the industrial denoising models and the state-of-the-art deep learning models on VoIP audio streams.

\section{Ablation Results}



\begin{table*}[!htp]\centering
\caption{Objective Relative Evaluation of Perceptual Quality \& Intelligibility}\label{tab:metrics_table_1}
\scriptsize
\sisetup{round-mode=places, round-precision=2}
\begin{tabular}{lrrrrrrrrrrrrrrrr}

\cellcolor[HTML]{d9d9d9} &\cellcolor[HTML]{d9d9d9} &\cellcolor[HTML]{d9d9d9}\textbar &\multicolumn{6}{c}{\cellcolor[HTML]{d9d9d9}PESQ} &\cellcolor[HTML]{d9d9d9}\textbar &\multicolumn{6}{c}{\cellcolor[HTML]{d9d9d9}STOI} \\ \hline

\cellcolor[HTML]{f3f3f3} &\cellcolor[HTML]{f3f3f3} &\cellcolor[HTML]{f3f3f3}\textbar &\multicolumn{2}{c}{\cellcolor[HTML]{f3f3f3}Source} &\multicolumn{2}{c}{\cellcolor[HTML]{f3f3f3}Demucs} &\multicolumn{2}{c}{\cellcolor[HTML]{f3f3f3}FullSubNet} &\cellcolor[HTML]{f3f3f3}\textbar &\multicolumn{2}{c}{\cellcolor[HTML]{f3f3f3}Source} &\multicolumn{2}{c}{\cellcolor[HTML]{f3f3f3}Demucs} &\multicolumn{2}{c}{\cellcolor[HTML]{f3f3f3}FullSubNet} \\
\hline

\cellcolor[HTML]{f3f3f3}Platform &\cellcolor[HTML]{f3f3f3}Receiver &\cellcolor[HTML]{f3f3f3}\textbar &\cellcolor[HTML]{f3f3f3}Low &\cellcolor[HTML]{f3f3f3}Auto &\cellcolor[HTML]{f3f3f3}Low &\cellcolor[HTML]{f3f3f3}Auto &\cellcolor[HTML]{f3f3f3}Low &\cellcolor[HTML]{f3f3f3}Auto &\cellcolor[HTML]{f3f3f3}\textbar &\cellcolor[HTML]{f3f3f3}Low &\cellcolor[HTML]{f3f3f3}Auto &\cellcolor[HTML]{f3f3f3}Low &\cellcolor[HTML]{f3f3f3}Auto &\cellcolor[HTML]{f3f3f3}Low &\cellcolor[HTML]{f3f3f3}Auto \\ 

Google Meets &Phone &\textbar & 1.549 & 1.976 & 1.381 & 1.726 & 1.398 & 1.694 & \textbar & 0.748 & 0.884 & 0.748 & 0.884 & 0.700 & 0.841 \\
Google Meets &Cloud &\textbar & 1.640 & 2.255 & 2.272 & 2.264 & 2.435 & 2.281 & \textbar & 0.890 & 0.923 & 0.933 & 0.924 & 0.920 & 0.923 \\
Zoom &Phone &\textbar & 1.548 & 1.701 & 1.548 & 1.513 & 1.382 & 1.343 & \textbar & 0.797 & 0.810 & 0.766 & 0.810 & 0.733 & 0.724 \\
Zoom &Cloud &\textbar & 1.450 & 1.919 & 2.089 & 1.996  & 2.141 & 1.992 & \textbar & 0.859 & 0.909 & 0.906 & 0.913 & 0.900 & 0.911 \\

\hline
\end{tabular}
\end{table*}

\begin{figure}[!htb]
    \centering
        \includegraphics[width=\linewidth]{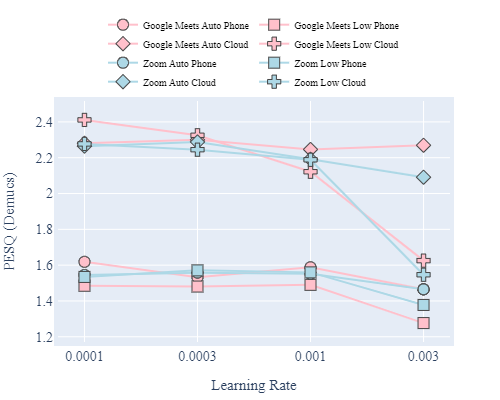}
        \caption{Ablation of Demucs  with different learning rate} 
        \label{fig:demucs_lr_ablation}
\end{figure}


We conducted a series of ablation experiments to evaluate the impact of our proposed temporal acoustic parameter (TAP) loss function on the denoising performance of our models. We trained with and without the TAP loss function, using different learning rates, and evaluated their performance on a large-scale dataset of 1,200 30-second noisy speech samples generated using the DNS-2020 synthesis procedure.

Our experimental results demonstrate that incorporating the TAP loss function leads to improved denoising performance over the baseline models. Specifically, we observed a notable improvement in the chosen metric for both Demucs and FullSubNet when incorporating the TAP loss function with an optimal learning rate. However, we also observed that the optimal learning rate for the TAP loss function varied depending on the specific characteristics of the VoIP platform and the transmission process.

In some cases, we found that a higher learning rate led to improved denoising performance on cellular devices when industrial denoising was disabled. This suggests that cellular devices may benefit from a more aggressive TAP loss function during training to achieve better denoising results. In contrast, for cloud relay transmissions and transmissions with industrial denoising enabled, a lower learning rate resulted in improved denoising performance.

Furthermore, we observed that the effectiveness of our proposed denoising models varied depending on the specific characteristics of the VoIP platform and the transmission process. Depending on the platform, whether or not industrial denoising was enabled, and which device received the audio, different acoustic distortions and artifacts could be heard.

In some scenarios, we observed that enhancing the audio without industrial denoising applied led to better denoising results. We believe that this is because less information is lost due to the platform's processing, which facilitates the restoration of the original audio. However, in other cases, we found that enabling industrial denoising led to better denoising results. This could be because some distortions and artifacts are magnified through transmission and are more present in audio without industrial denoising. In these cases, enabling industrial denoising may remove some information from the audio, but the benefits of removing much of the distortions and artifacts prior to transmission outweigh the cons.

Taken together, our results demonstrate that the effectiveness of denoising models on VoIP platforms is highly dependent on the specific characteristics of the platform, the transmission process, and the optimal learning rate used during training. Nonetheless, our proposed denoising models consistently outperformed the industrial denoising models when industrial denoising was disabled, suggesting that our proposed models have the potential for improving denoising performance on VoIP platforms.

\begin{figure*}[!htb]
    \centering
    \begin{subfigure}{0.5\textwidth}
        \includegraphics[width=\linewidth]{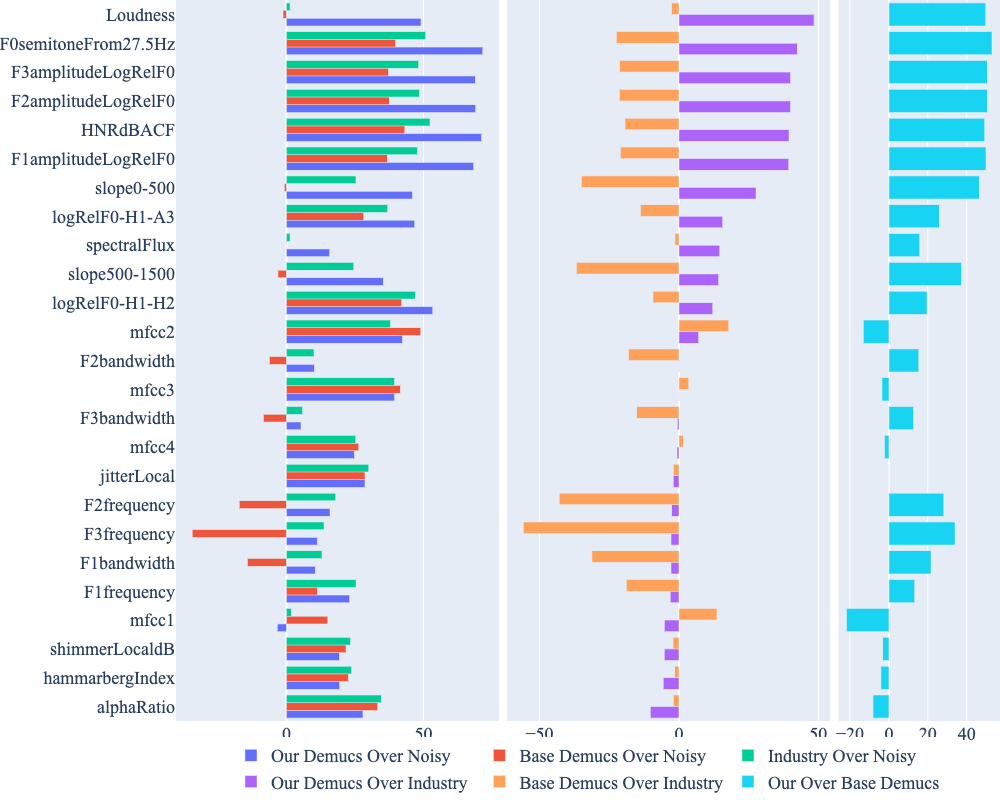}
        \caption{Demucs Standardized Acoustic Improvement Comparison}
        \label{fig:demucs_acoustics}
    \end{subfigure}%
    \begin{subfigure}{0.5\textwidth}
        \includegraphics[width=\linewidth]{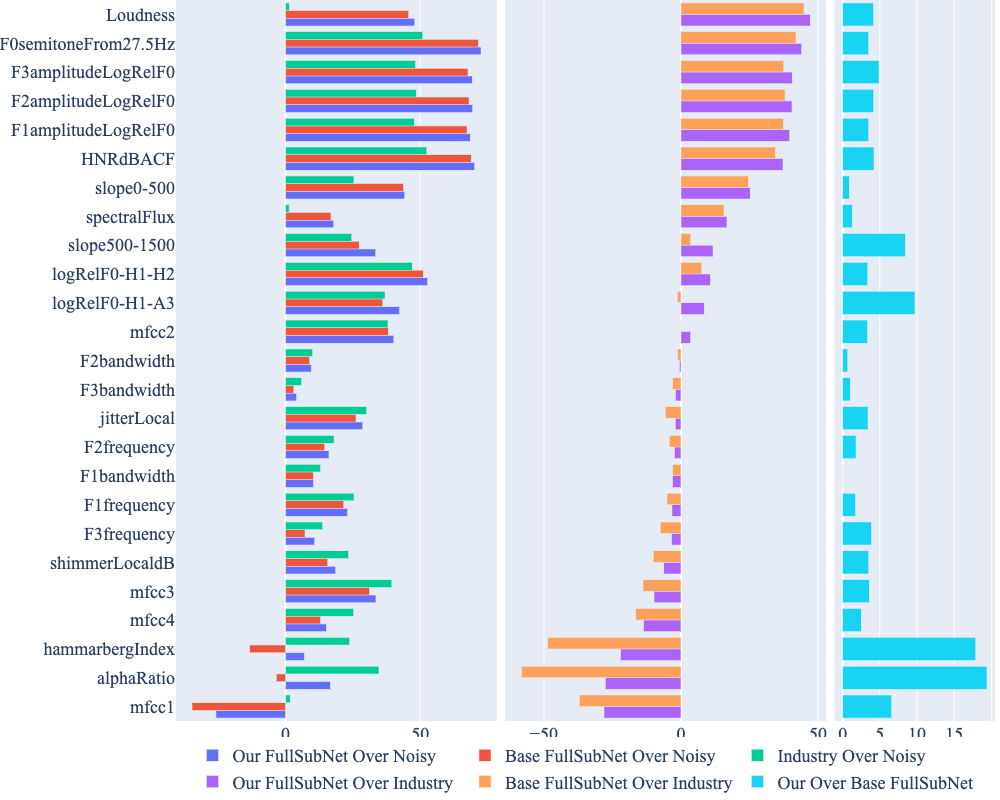}
        \caption{FullSubNet Standardized Acoustic Improvement Comparison}
        \label{fig:fsn_acoustics}
    \end{subfigure}
    \caption{ Acoustic Improvement Of Mean Absolute Error: The Use Case Of Transmitting From Google Meet To Cellular Phone }
    \label{fig:acoustics_improv}
\end{figure*}

\section{Perceptual Evaluation}

In this section, we present a perceptual analysis of our fine-tuning results, investigating intelligibility and perceptual quality of the models on different VoIP applications. To assess intelligibility, we use the Short-Time Objective Intelligibility (STOI)\cite{stoi} measure, which quantifies the extent to which speech can be understood by human listeners. For evaluating perceptual quality, we employ the Perceptual Evaluation of Speech Quality (PESQ)\cite{pesq} metric, which approximates human subjective judgments of speech quality. We provide a subsection for the overall comparison, Demucs vs FullSubNet, and differential analysis of low vs auto specifications.

\subsection{Overall Comparison}

Our fine-tuning approach was able to match or exceed industry intelligibility levels. While cloud applications exhibited little difference, cellular phone applications showed a 0.7\% STOI improvement. Additionally, we observed perceptual quality improvements of 1.2\% on Google Meet and 4.0\% on Zoom over industry performance for cloud applications. However, mobile applications demonstrated worse perceptual performance than industry standards, which is likely attributed to the challenge of transfer learning from DNS-2020 to the band limitations of cellular transmission.

\subsection{Demucs vs FullSubNet Comparison}

Upon comparing Demucs and FullSubNet, we found that Demucs had superior intelligibility, with a PESQ improvement ranging from 0.2-1.1\% on cellular applications to 5.1-10.5\% on cloud applications. The perceptual improvement depended on the receiving medium, with Demucs outperforming FullSubNet by 1.8-12.0\% on mobile applications and FullSubNet surpassing Demucs by 2.5-7.2\% on cloud applications. Although a detailed error analysis is beyond the scope of this paper, we observed that the band limitations of modeling the time-frequency domain caused a noticeable distribution shift, making transfer learning on cellular applications slower to converge. Gathering more data, especially for cellular applications, could help overcome this challenge.

\subsection{Denoising Off (Low) vs Denoising On (Auto) Settings}

To better understand the relative performance specific to relays without denoising prior to transmission (low) versus relays with denoising prior to transmission (auto), we conducted a differential analysis.

For the case without denoising prior to transmission, we observed superior perceptual quality with modest PESQ improvement across applications, except for Google Meets to a cellular phone. In contrast to Zoom, Google Meets experienced worse perceptual quality due to noisy audio transmission, with a PESQ difference of about 0.14, which could explain the difficulty in learning to enhance the speech signal.

Comparing within Demucs (auto vs low), relays with denoising prior to transmission yielded comparable or improved intelligibility over industry standards, with STOI improvement observed when transmission was received from Google Meets to a cellular phone (18.2\%), Zoom to a cellular phone (5.7\%), and to some extent from Zoom to the cloud (0.7\%). However, transmission from Google Meets to  cloud witnessed a modest STOI degradation of 1.0\%, making low performance superior to auto. We conjecture that specifically for the Google Meets to Cloud use case, the exacerbation of distortion or artifact issues in the low use case was not as significant as the amount of information lost due to their denoising prior to transmission. These nuances and complexities highlight the importance of platform-specific assessment, as the areas in need of improvement and their relative difficulty to improve can vary.

In conclusion, our perceptual analysis demonstrates the potential benefits and challenges of applying our fine-tuning approach to different VoIP platforms and scenarios. The results emphasize the importance of platform-specific assessment to optimize factors impacting speech enhancement performance.

\section{Acoustic Evaluation}

In this section, we present an acoustic analysis of the denoising models, focusing on the improvement of fine-grain speech characteristics over noisy speech, industrial denoising models of Zoom and Google Meet, and the baseline models. We use the extended Geniva Minimalist Acoustic Parameter Set (eGeMAPS) \cite{eyben2015geneva} for this evaluation.

For each test sample, we extracted eGeMAPS features from the original clean speech, the noisy speech, and the denoised speech generated by our proposed models, the industrial models, and the baseline models. We then calculated the Mean Absolute Error (MAE) between the eGeMAPS features of the original clean speech and the degraded and enhanced versions of the speech. Due to space constraints, we focus on Google Meet to Phone acoustic improvement for our approach tuning Demucs and FullSubNet compared to the industry and their relative improvement over their open-source baselines.

In almost all scenarios, the acoustics of fine-grain speech characteristics improved through denoising when applied only before transmission, only after transmission, and both before and after transmission. Our approach led to mixed results in achieving industry performance, but when compared to the open-source alternative, tailoring Demucs and FullSubNet on DNS-2020 to the VoIP-DNS set yielded as high as 15\% improvement for FullSubNet and 40\% for Demucs, with an average improvement in acoustic fidelity of about 5\% for FullSubNet and 21\% for Demucs.

Compared to industry performance, we achieved higher acoustic fidelity in amplitudeLogRelF0, with Demucs showing approximately 40\% improvement over the industry and 51\% improvement over the baseline. FullSubNet demonstrated a modest 4\% improvement over the baseline with 40\% improvement over industry across formants F1, F2, and F3. However, in formant-specific bandwidth improvement, the difference between Demucs or FullSubNet and industry was marginal.

Other acoustics that showed significant improvement over industry performance were Loudness (48\% for Demucs and 47\% for FullSubNet) and Harmonic to Noise Ratio Autocorrelation Function (HNRdbACF) (39\% for Demucs and 37\% for FullSubNet), suggesting proper normalization and ability discriminate between harmonic and non-harmonic signals in our approach. While further analysis is required to understand the relationship between these acoustic changes and perceptual quality, reducing harmonic distortion and improving loudness accuracy may contribute to better subjective evaluation.

In some areas, such as MFCCs, the Hammarberg Index, and the Alpha Index, Demucs and FullSubNet had difficulty improving acoustics. However, our approach experienced less degradation in these areas compared to the baseline. We believe further improvement is possible, and we are currently conducting ablations to refine our models. Updated findings reflecting these insights will be reported in the near future.

\section{Conclusion}

In this study, we have investigated the denoising performance of two state-of-the-art deep learning models, Demucs and FullSubNet, on VoIP audio streams, with a particular focus on their application in popular platforms such as Zoom and Google Meet. By fine-tuning these models on the VoIP-DNS dataset and incorporating a novel acoustic loss function, we achieved significant improvements in denoising performance compared to their open-source baselines.

Our perceptual analysis demonstrated that our fine-tuned models were able to match or exceed the intelligibility of industry-denoising models in various scenarios, with Demucs generally outperforming FullSubNet in terms of PESQ and STOI. The differential analysis of low vs auto specifications revealed that our models performed differently depending on the relay condition and the platform used, emphasizing the importance of platform-specific assessment and optimization. We also observed differences in performance between phone and cloud applications, with cloud applications generally exhibiting better perceptual quality improvements.

The acoustic analysis using eGeMAPS features further supported the effectiveness of our approach in improving fine-grain speech characteristics. Our models demonstrated substantial improvements in certain acoustic features, such as amplitudeLogRelF0, Loudness, and HNRdBACF, compared to the industry and open-source baselines.

However, there are still areas for further improvement, such as in MFCCs, the Hammarberg Index, and the Alpha Index. We believe that additional fine-tuning and ablation studies will help address these issues and lead to even better denoising performance in the future.

This study contributes to the growing body of research on speech enhancement for VoIP applications and highlights the potential of deep learning-based denoising models to improve the quality of real-time communication. As the demand for effective remote communication continues to grow, the advancements presented in this work serve as a foundation for further development and optimization of speech enhancement techniques tailored to the specific requirements of various VoIP platforms, transmission scenarios, and end-user devices, including both phone and cloud applications.

\bibliographystyle{IEEEtran}

\bibliography{template}

\appendix

\section{Appendix}
\subsection{The Temporal Acoustic Parameter Loss (TAPLoss)}
\label{appendix-taploss}
To calculate the TAPLoss, we define the Temporal Acoustic Parameter Estimator $A_y$. $A_y(t, p)$ represents the acoustic parameter $p$ at a discrete time frame $t$. To predict $A_y$, we define the estimator:
$$\hat{A_y} = \Tc\Ac\Pc(y)$$

The $\Tc\Ac\Pc$ function takes in a signal input $y$. It calculates a complex spectrogram $Y$ with $F = 257$ frequency bins. It then passes this complex spectrogram to a recurrent neural network to output the temporal acoustic parameter $\hat{A_y}$. TAP loss is then defined as the mean average error between the actual and the predicted estimate.

$$MAE(A_y, \hat{A_y}) = \frac{1}{TP} \sum_{t=0}^{T-1} \sum_{p=0}^{P-1} | A_y(t, p) - A_{\hat{y}}(t, p) | \in R$$

During training, $\Tc\Ac\Pc$ parameters learn to minimize the divergence of $MAE(A_s, \hat{A_s})$ using Adam optimization. Since this loss is end-to-end differentiable and takes only waveform as input, it enables acoustic optimization of any speech model and task with clean references. 

    
    
    

\clearpage

\end{document}